\documentclass[a4paper]{jpconf}
\usepackage{graphicx}
\usepackage{float}
\usepackage{amsmath}
\usepackage{amssymb}
\usepackage{multirow}

\usepackage{epsfig}
\graphicspath {{./figures/}}
\usepackage{color,soul}

\begin{document}

\title{Corrections to scaling in geometrical clusters of the 2D Ising model}

\author{Michail Akritidis$^1$, Nikolaos G Fytas$^1$, and Martin Weigel$^{2\;1}$}

\address{$^1$ Centre for Fluid and Complex Systems, Coventry
	University, Coventry, CV1 5FB, United Kingdom}

\address{$^2$ Institut f\"{u}r Physik, Technische Universit\"{a}t Chemnitz, 09107 Chemnitz, Germany}

\ead{akritidm@uni.coventry.ac.uk}


\begin{abstract}
	
We study the scaling of the average cluster size and percolation  strength of geometrical clusters for the two-dimensional Ising model. By means of Monte Carlo simulations and a finite-size scaling analysis we discuss the appearance of corrections to scaling for different definitions of cluster sets. We find that including all percolating clusters, or excluding only clusters that percolate in one but not the other direction, leads to smaller corrections to scaling for the average cluster size as compared to the other definitions considered. The percolation strength is less sensitive to the definitions used.

\end{abstract}


\section{Introduction} \label{sec:introduction}


Percolation theory \cite{stauffer_introduction_1994} is closely linked to critical phenomena in spin models, as it provides geometrical insights regarding the nature of the transition. The percolation problem is defined through clusters of occupied sites or bonds on a lattice, and one studies the geometrical transition at the percolation threshold, characterized by an incipient spanning cluster that becomes infinite in the thermodynamic limit. The $ q $-state Potts model can be mapped onto a percolation process, by constructing clusters of neighbouring spins with the same orientation. Unfortunately, these \textit{geometrical} clusters do not in general percolate at the critical temperature $ T_{\text{c}} $ of the actual phase transition, and they hence do not properly describe the thermal correlations of the system (see \cite{coniglio_correlated_2009, saberi_recent_2015} for a review). Using a different cluster definition where bonds of neighbouring like spins are deleted with a certain temperature-dependent probability, Fortuin and Kasteleyn (FK) \cite{fortuin_random-cluster_1972} showed that the model can be mapped onto a modified percolation process, where clusters percolate at $ T_{\text{c}}$ and, more importantly, they encode the critical behaviour of the system, resulting in the same set of critical exponents. The FK clusters are also the core idea behind the powerful Monte Carlo algorithms of Swendsen and Wang \cite{swendsen_nonuniversal_1987} and Wolff \cite{wolff_collective_1989}, where entire FK clusters are flipped instead of individual spins as in the standard Metropolis algorithm  \cite{metropolis_equation_1953}. The main advantage of these non-local cluster updates over the local spin flips is the reduction of autocorrelation times in the vicinity of the critical point, significantly improving the simulation efficiency.

Although geometrical clusters do not capture the critical behaviour, in the two-dimensional (2D) Potts model they at least do percolate at $ T_{\text{c}} $. It turns out that for $ 0\le q \le 4 $ the critical behaviour of geometrical clusters for the pure model is related to the tricritical behaviour of a diluted Potts model \cite{janke_geometrical_2004, janke_geometric_2006}. Stella and Vanderzande \cite{stella_vanderzande_scaling_1989,vanderzande_stella_bulk_1989}
showed that the geometrical clusters of the $ q=2 $ Potts model, i.e., the Ising model, follow the tricritical behaviour of the $ q=1 $ site diluted Potts model. This aspect was numerically verified in references \cite{janke_geometrical_2004, janke_geometric_2006}.

In the present study we investigate numerically the critical behaviour of the geometrical clusters of the square-lattice Ising model, whose Hamiltonian reads

\begin{equation}\label{eq:Ising Hamiltonian}
	\mathcal{H} = -J\sum_{\langle ij \rangle}S_{i}S_{j},
\end{equation}

\noindent
where the spins take on the values $\pm 1$, $J>0$ is the ferromagnetic exchange interaction, and the sum extends over all nearest neighbours of the lattice. Specifically, we estimate the exponents of the average cluster size and percolation strength for different sets of clusters, while monitoring the existence of corrections to scaling in each case.



\section{Observables} \label{sec:observables}

As in random percolation \cite{stauffer_introduction_1994}, the quantities of interest are the percolation strength $ P_{\infty} $, which gives the fraction of sites belonging to the infinite cluster in the thermodynamic limit, and the average cluster size

\begin{equation}\label{eq:aver.clust.size}
	S = \frac{\sum_{s'} s^2 \; n_s}{\sum_{s'} s \; n_s},	
\end{equation}

\noindent
where $ s $ denotes the size (number of spins) of the cluster and $ n_s $ the number of clusters with size $ s $. The notation $ s' $ indicates that the sums are restricted to different sub-sets of clusters, which for the present study are: 

\begin{enumerate}\setlength\itemsep{-0.0001em}
	\item All clusters are included: $ C $.
	\item Exclude the largest cluster in each measurement: $C \; \backslash \; \text{max} \; C$.
	\item Exclude all percolating clusters: $C \; \backslash \; P$.
	\item Exclude all clusters percolating in horizontal and in vertical direction: $C \; \backslash \; P_{\text{x and y}}$.
	\item Exclude all clusters percolating in one specific direction, e.g., horizontal: $C \; \backslash \; P_{\text{x}}$. 
	\item Exclude all clusters percolating in one but not the other direction, e.g., horizontal and not vertical: $C \; \backslash \; P_{\text{x and } \overline{\text{y}}}$.
\end{enumerate}

For finite size systems $ P_{\infty} $ is usually estimated by the fraction of sites belonging to the largest cluster. Following the idea of cluster sets for $ S $, we can introduce similar sets for $ P_{\infty} $. In the present study we consider, in each configuration, the fraction of sites that belong to the:

\begin{enumerate}\setlength\itemsep{-0.0001em}
	\item Largest cluster: $\text{max} \; C$.
	\item Largest percolating cluster: $\text{max} \; P$.
	\item Largest cluster that percolates in horizontal and in vertical direction: $\text{max} \; P_{\text{x and y}}$.
	\item Largest cluster that percolates in one specific direction, e.g., horizontal: $\text{max} \; P_{\text{x}}$.
	\item Largest cluster that percolates in one but not the other direction, e.g., horizontal and not vertical: $\text{max} \; P_{\text{x and } \overline{\text{y}}}$.
\end{enumerate}

\noindent
In the vicinity of the critical point these observables exhibit a scaling behaviour of the form \cite{stauffer_introduction_1994} 

\begin{align}
	\label{eq:strength scaled}
	P_{\infty}\left(L,T\right) &= L^{-\beta / \nu} f_{P_{\infty}}\left[\left(T-T_{\text{c}}\right) L^{1/ \nu}\right], \\
	\label{eq:aver.clust.size scaled}
	S\left(L,T\right) &= L^{\gamma / \nu} f_{S}\left[\left(T-T_{\text{c}}\right) L^{1/ \nu}\right],  
\end{align} 
\noindent
where $\nu$ is the critical exponent of the correlation length, $\gamma$ the critical exponent that captures the divergence of the average cluster size, and $\beta$ the critical exponent which describes how the strength of the percolating cluster goes to zero as $T_{\rm c}$ is approached from below. $ f_{P_{\infty}} $ and $ f_{S} $ are universal finite-size scaling functions.


\section{Results} \label{sec:results}

We simulated the 2D  Ising model with periodic boundary conditions using the Swendsen-Wang algorithm \cite{swendsen_nonuniversal_1987}. We considered systems of linear size  $L = 16,$  $ 32,$  $ 64,$  $128,$  $256,$  $512,$  $600,$  $1000,$  $1200,$  $1600$  $ \text{and } 2000$, at the exact critical temperature $T_{\rm c} = 2/\ln\left(1+\sqrt{2}\right)$. For all $ L $ the total number of simulation steps was $ 1.1\times \tau_{\text{int, E}} \times 10^5 $ sweeps, where $ \tau_{\text{int, E}} $ is the integrated autocorrelation time of the energy \cite{weigel_error_2010}, and $\tau_{\text{int, E}} \times 10^4$ sweeps were discarded during equilibration. After every $ \tau_{\text{int, E}} $ sweeps a measurement was taken, leading to up to $ 10^5 $ measurements per run. The estimates of $ \tau_{\text{int, E}} $, rounded up to the next largest integer, in ascending order of the system size are: $ \tau_{\text{int, E}} = $ $ 4,$  $ 5,$  $ 5,$  $ 6,$  $ 7,$  $ 9,$  $ 9,$  $ 10,$  $ 11,$  $ 11,$  $ 12 $ sweeps. Statistical errors were estimated by means of jackknife blocking \cite{efron_introduction_1994}. We considered a cluster to percolate in one direction if it wraps around this direction and is connected back to itself. To identify the wrapping clusters, we employed the method of Machta \textit{et al.} \cite{machta_invaded_1996}.


At $ T_{\text{c}} $, according to equations (\ref{eq:strength scaled}) and (\ref{eq:aver.clust.size scaled}),  $ f_P(0) = \text{const}. $ and $ f_S(0) = \text{const.}$, allowing the estimation of $\gamma / \nu $ and $ \beta / \nu $ as a function of $ L $. For all sets of clusters, fits were performed using the least-squares Levenberg-Marquardt algorithm \cite{press_etal_1992}. In order to obtain the influence of corrections to scaling in the observables we performed fits including $ L_{\text{min}} \le L \le L_{\text{max}} $, by systematically increasing the lower cut-off $ L_{\text{min}} $, while the upper cut-off is kept fixed $ L_{\text{max}} = 2000 $.

\begin{figure}
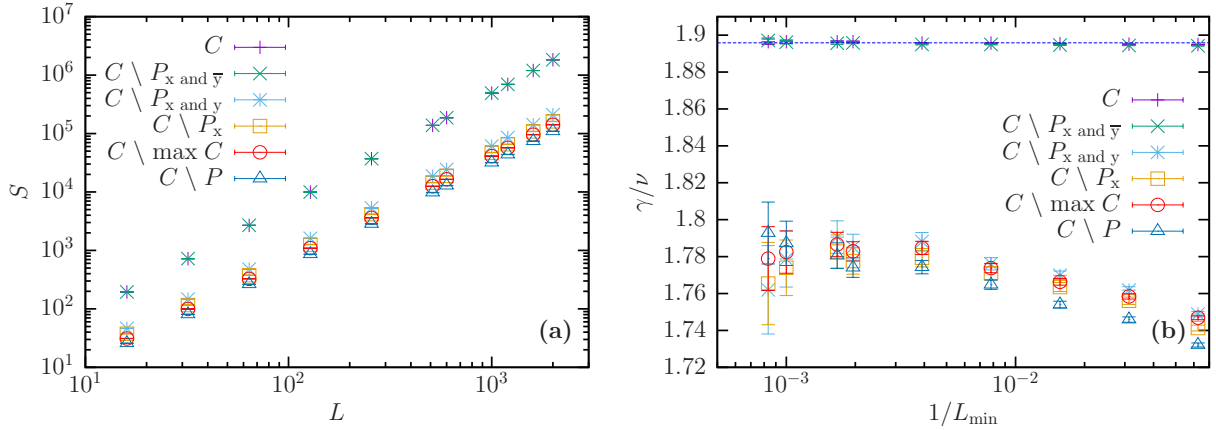

	\centering
	\begin{minipage}[t]{0.45\linewidth}
		\hspace*{-0.5cm}
		\resizebox{1.13\linewidth}{!}{{\large \input{Figures/cluster_size_vs_L_q1-different_definitions.tex}}}
	\end{minipage}
	\hspace{.2in}	
	\begin{minipage}[t]{0.45\linewidth}
		\hspace*{-0.3cm}\centering
		\resizebox{1.13\linewidth}{!}{{\large \input{Figures/gamma_vs_1_over_Lmin_q1-different_definitions.tex}}}
	\end{minipage}
	\caption{(a) Log-log plot of the average cluster size as a function of $ L $ for the different definitions. (b) Exponent ratio $ \gamma / \nu $ as a function of $ 1 / L_{\text{min}} $. The dashed horizontal line marks the asymptotic value $ 91/48 \approx 1.896 $  \cite{stella_vanderzande_scaling_1989}.}
	\label{fig:average cluster size}
\end{figure}

In figure \ref{fig:average cluster size}(a) the average cluster size as a function of $ L $ is plotted for the different definitions introduced in section \ref{sec:observables}. Visually, the data of all definitions seem to follow straight lines, on a log-log plot, indicating merely small corrections to scaling. Estimates of $ \gamma / \nu $ for different fit intervals and the respective quality-of-fit parameter $ Q $  are reported in table \ref{table:gamma_different_definitions-q1}. For $ C $ and $C \; \backslash \; P_{\text{x and } \overline{\text{y}}}$, $ \gamma / \nu $ converges relatively quickly to its asymptotic value, $ \gamma / \nu =  91 / 48 \approx 1.896 $ \cite{stella_vanderzande_scaling_1989} as $ L_{\text{min}} $ increases. Specifically, the discrepancy from the asymptotic value is less than 3 standard deviations $ \left( 3\sigma \right) $ for $ L_{\text{min}} \ge 64 $ and $ L_{\text{min}} \ge 256 $ respectively, so that corrections to scaling are not substantial. This is nicely illustrated in figure 1(b), where
the exponent $ \gamma / \nu $ is plotted as a function of $ 1 / L_{\text{min}} $. For $C \; \backslash \; P_{\text{x and y}}$, $C \; \backslash \; P_{\text{x}}$, $C \; \backslash \; \text{max} \; C$, and $C \; \backslash \; P$, on the other hand, the convergence to the asymptotic value is  found to be very slow. Even for $ L_{\text{min}} = 1200 $, the deviation from the asymptotic values is around  $  6\sigma $ for all definitions, indicating that corrections to scaling are important. This is illustrated in figure \ref{fig:average cluster size}(b), where the estimates of the exponent seem to converge to a value significantly below the predicted one. The fact that $ C $ and $C \; \backslash \; P_{\text{x and } \overline{\text{y}}}$  give similar results is to be expected. Percolating clusters in one but not the other direction are very rare, thus excluding them from the sums of equation (\ref{eq:aver.clust.size}) will not affect the average cluster size significantly. This can also be seen from figure  \ref{fig:average cluster size}(a), where the data points for the two definitions, at the same $ L $, are equal within error bars. Note that previous estimates of the exponent using the $ C $ definition are reported in reference \cite{janke_fractal_2005} and these for the $C \; \backslash \; P_{\text{x}}$ definition in reference \cite{fortunato_site_2002}.  

\begin{table}
	\caption{\label{table:gamma_different_definitions-q1} Estimates of $  \gamma / \nu$, quality-of-fit parameter $ Q $, and deviations from the exact value, $ \gamma / \nu = 91/48 \approx 1.896 $, in multiples of the estimated statistical errors $ \Delta\left(\sigma\right) $, for the cluster sets of the average cluster size as introduced in section \ref{sec:observables}.}
	\begin{center}
		\begin{tabular}{llllllllll}
			\br
			&\multicolumn{3}{l}{$ C $} &\multicolumn{3}{l}{$C \; \backslash \; P_{\text{x and } \overline{\text{y}}}$}& \multicolumn{3}{l}{$C \; \backslash \; P_{\text{x and y}}$}\\
			\mr
			$L_{\text{min}}$&$\gamma/\nu$&$Q$&$ \Delta\left(\sigma\right) $&$\gamma/\nu$&$Q$&$ \Delta\left(\sigma\right) $&$\gamma/\nu$&$Q$&$ \Delta\left(\sigma\right) $ \\
			\mr
			
			16 & 1.89490(11) & 0.001 & 8.28   &
			1.89419(11) & 0.001 & 15.03  &
			1.7488(14)  & 0.000 & 101.59 \\
			
			32 & 1.89510(14) & 0.010 & 5.36  &
			1.89440(13) & 0.008 & 10.76 &
			1.7618(19)  & 0.000 & 71.51 \\
			
			64 & 1.89535(17) & 0.044 & 2.83  &
			1.89465(17) & 0.042 & 7.11  &
			1.770(2)    & 0.002 & 53.39 \\
			
			128 & 1.8957(2) & 0.123 & 0.75  &
			1.8950(2) & 0.130 & 3.94  &
			1.776(3)  & 0.021 & 36.14 \\
			
			256 & 1.8957(3) & 0.078 & 0.52  &
			1.8950(3) & 0.079 & 2.78  &
			1.788(5)  & 0.658  & 22.21 \\
			
			512 & 1.8964(4) & 0.723 & 1.30  &
			1.8958(4) & 0.747 & 0.10  &
			1.785(7)  & 0.575 & 15.78 \\
			
			600 & 1.8964(6) & 0.613 & 1.03  &
			1.8956(6) & 0.668 & 0.46  &
			1.791(9)  & 0.611 & 11.97 \\
			
			1000 & 1.8964(10) & 0.568 & 0.60 &
			1.8962(9)  & 0.623 & 0.35 &
			1.779(15)  & 0.639 & 7.71 \\
			
			1200 & 1.8964(14) & 0.300 & 0.40 &
			1.8971(14) & 0.628 & 0.86 &
			1.76(2)    & 0.778 & 5.59 \\ \\

			&\multicolumn{3}{l}{$C \; \backslash \; P_{\text{x}}$} &\multicolumn{3}{l}{$C \; \backslash \; \text{max} \; C$}& \multicolumn{3}{l}{$C \; \backslash \; P$}\\
			\mr
			$L_{\text{min}}$&$\gamma/\nu$&$Q$&$ \Delta\left(\sigma\right) $&$\gamma/\nu$&$Q$&$ \Delta\left(\sigma\right) $&$\gamma/\nu$&$Q$&$ \Delta\left(\sigma\right) $ \\
			\mr    
			
			16 & 1.7413(16) & 0.000 & 99.00  &
			1.7469(11) & 0.000 & 140.19 &
			1.7321(11) & 0.000 & 148.24 \\
			
			32 & 1.756(2)   & 0.000 & 70.11  &
			1.7584(14) & 0.000 & 97.21  &
			1.7459(15) & 0.000 & 102.89 \\
			
			64 & 1.764(2)    & 0.016 & 54.90 &
			1.7663(18)  & 0.000 & 70.59 &
			1.7540(18) & 0.000 & 76.73 \\
			
			128 & 1.771(3) & 0.307 & 37.30 &
			1.774(2) & 0.003 & 48.41 &
			1.765(3) & 0.008 & 52.35 \\

			256 & 1.779(5) & 0.875 & 23.45 &
			1.785(4) & 0.868 & 30.87 &
			1.774(4) & 0.514 & 33.64 \\
			
			512 & 1.778(7) & 0.794 & 16.59 &
			1.783(5) & 0.806 & 21.64 &
			1.774(5) & 0.374 & 23.01 \\
			
			600 & 1.783(9) & 0.847 & 12.52 &
			1.787(7) & 0.860 & 16.57 &
			1.781(7) & 0.544 & 16.65 \\
			
			1000 & 1.774(15) & 0.873 & 8.11 &
			1.782(11) & 0.754 & 9.87 &
			1.787(12) & 0.431 & 9.04 \\
			
			1200 & 1.77(2)   & 0.969 & 5.85 &
			1.779(17) & 0.484 & 6.76 &
			1.793(17) & 0.228 & 6.15 \\
			
			\br
		\end{tabular}
	\end{center}	
\end{table}

\begin{figure}[]
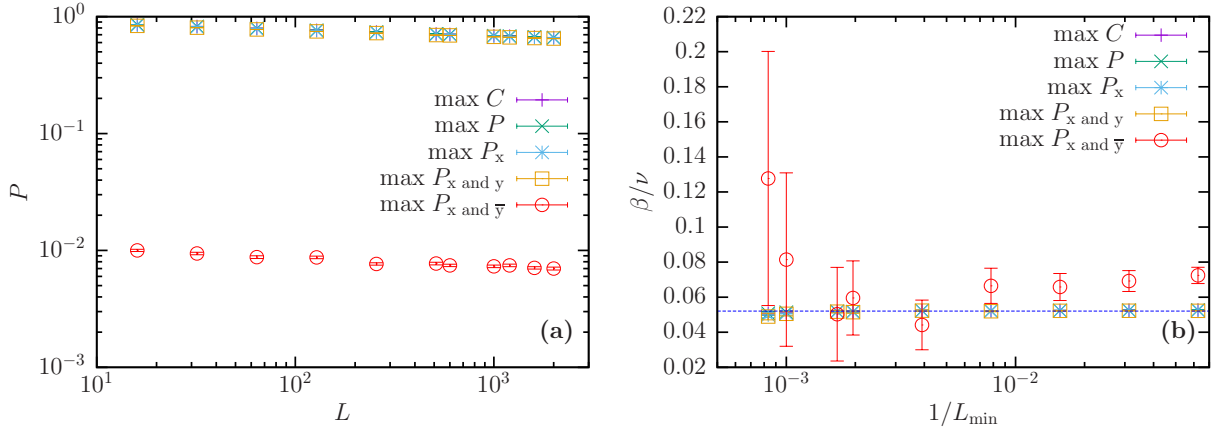

	\centering
	\begin{minipage}[t]{0.45\linewidth}
		\hspace*{-0.5cm}
		\resizebox{1.13\linewidth}{!}{\large {\input{Figures/strength_vs_L_q1-different_definitions.tex}}}
	\end{minipage}
	\hspace{.2in}	
	\begin{minipage}[t]{0.45\linewidth}
		\hspace*{-0.3cm}\centering
		\resizebox{1.13\linewidth}{!}{\large {\input{Figures/beta_vs_1_over_Lmin_q1-different_definitions.tex}}}
	\end{minipage}
	\caption{(a) Log-log plot of the strength as a function of $ L $ for the different definitions. (b) Exponent ratio $ \beta / \nu $ as a function of $ 1 / L_{\text{min}} $. The dashed horizontal line marks the asymptotic value $ 5/96 \approx 0.052 $  \cite{stella_vanderzande_scaling_1989}.}
	\label{fig:strength}
\end{figure}

For the percolation strength, data for all sets seem to follow a straight line on a log-log plot as a function of $ L $, indicating small corrections to scaling, as is shown in figure \ref{fig:strength}(a). This observation is supported by table \ref{table:beta_different_definitions-q1}, where the estimates of $ \beta / \nu $ and $ Q $ are reported. Setting aside $\text{max} \; P_{\text{x and } \overline{\text{y}}}$, for all sets the deviations from the asymptotic value $ \beta / \nu = 5/96 \approx 0.052 $ \cite{stella_vanderzande_scaling_1989} is less than $ 3 \sigma $ for all $ L_{\text{min}} \ge 64 $, indicating rather small corrections to scaling. Notably, for $\text{max} \; P_{\text{x and y}}$ deviations are even smaller  than $ 1.7\sigma $ for all $ L_{\text{min}} $. According to table \ref{table:beta_different_definitions-q1} the estimates of $ \beta / \nu $ for $\text{max} \; P_{\text{x and } \overline{\text{y}}}$ are less than $ 3 \sigma $ from the asymptotic value (except $ L_{\text{min}} = 16 $), but the statistical errors are large, thus not allowing a reasonable estimate of the exponent. The explanation for this behaviour of $\text{max} \; P_{\text{x and } \overline{\text{y}}}$ is in line with the $C \; \backslash \; P_{\text{x and } \overline{\text{y}}}$ for the average cluster size, but going in the opposite direction. As already stated above, clusters percolating in one but not the other direction are rare, leading to poor statistics and larger statistical errors for the percolation strength; see figure \ref{fig:strength}(b), where the exponent $ \beta / \nu $ is plotted as a function of $ 1 / L_{\text{min}} $.

\begin{table}[]
	\caption{\label{table:beta_different_definitions-q1} Estimates of $  \beta / \nu$, quality-of-fit parameter $ Q $, and deviations from the exact value, $ \beta / \nu = 5/96 \approx 0.052 $, in multiples of the estimated statistical errors $ \Delta\left(\sigma\right) $, for the cluster sets of the percolation strength as introduced in section \ref{sec:observables}.}
	\begin{center}
		\begin{tabular}{llllllllll}
			\br
			&\multicolumn{3}{l}{$\text{max} \; C$} &\multicolumn{3}{l}{$\text{max} \; P$}& \multicolumn{3}{l}{$\text{max} \; P_{\text{x}}$}\\
			\mr
			$L_{\text{min}}$&$\beta/\nu$&$Q$&$ \Delta\left(\sigma\right) $&$\beta/\nu$&$Q$&$ \Delta\left(\sigma\right) $&$\beta/\nu$&$Q$&$ \Delta\left(\sigma\right) $ \\
			\mr
			
			16 & 0.05271(8)  & 0.005 & 7.94 &
			0.05269(8)  & 0.008 & 7.65 &
			0.05245(11) & 0.230 & 3.30 \\
			
			32 & 0.05256(10) & 0.030 & 4.94 &
			0.05255(10) & 0.036 & 4.76 &
			0.05243(14) & 0.166 & 2.56 \\
			
			64 & 0.05241(12) & 0.081 & 2.66 &
			0.05240(12) & 0.092 & 2.55 &
			0.05236(16) & 0.136 & 1.68 \\
			
			128 & 0.05222(17) & 0.141 & 0.81 &
			0.05222(17) & 0.136 & 0.83 &
			0.0521(2)   & 0.216 & 0.06 \\
			
			256 & 0.0523(2) & 0.109 & 1.14 &
			0.0523(2) & 0.103 & 1.13 &
			0.0522(3) & 0.156 & 0.43 \\
			
			512 & 0.0517(3) & 0.744 & 1.06 &
			0.0517(3) & 0.718 & 1.07 &
			0.0515(5) & 0.597 & 1.32 \\
			
			600 & 0.0519(4) & 0.670 & 0.44 &
			0.0519(4) & 0.653 & 0.42 &
			0.0517(6) & 0.494 & 0.70 \\
			
			1000 & 0.0514(7) & 0.671 & 0.96 &
			0.0514(7) & 0.678 & 0.99 &
			0.0507(9) & 0.681 & 1.45 \\
			
			1200 & 0.0509(11) & 0.586 & 1.17 &
			0.0508(11) & 0.600 & 1.19 &
			0.0498(14) & 0.955 & 1.61 \\ \\

			&\multicolumn{3}{l}{$\text{max} \; P_{\text{x and y}}$} &\multicolumn{3}{l}{$\text{max} \; P_{\text{x and } \overline{\text{y}}}$}&\multicolumn{3}{l}{ }\\
			\mr
			$L_{\text{min}}$&$\beta/\nu$&$Q$&$ \Delta\left(\sigma\right) $&$\beta/\nu$&$Q$&$ \Delta\left(\sigma\right) $& - & - & - \\
			\mr    
			
			16 & 0.05224(14) & 0.232 & 1.17 &
			0.072(5)    & 0.432 & 4.38 \\
			
			32 & 0.05227(17) & 0.169 & 1.08 &
			0.069(6)    & 0.402 & 2.87 \\
			
			64 & 0.0522(2) & 0.119 & 0.66 &
			0.066(8)  & 0.347 & 1.78 \\
			
			128 & 0.0520(3) & 0.135 & 0.42 &
			0.066(10) & 0.251 & 1.43 \\
			
			256 & 0.0523(4) & 0.153 & 0.60 &
			0.044(14) & 0.730 & 0.56 \\
			
			512 & 0.0514(6) & 0.508 & 1.14 &
			0.06(2)   & 0.763 & 0.35 \\
			
			600 & 0.0517(7) & 0.435 & 0.48 &
			0.05(3)   & 0.676 & 0.07 \\
			
			1000 & 0.0505(12) & 0.569 & 1.30 &
			0.08(5)    & 0.617 & 0.59 \\	
			
			1200 & 0.0490(19) & 0.881 & 1.64 &
			0.13(7)    & 0.651 & 1.04 \\ 	
			
			\br
		\end{tabular}
	\end{center}	
\end{table}


\section{Summary}\label{sec:summary}


Using several classes of sub-sets of clusters we determined the critical exponents for the average cluster size and percolation strength for geometrical clusters in the 2D Ising model. Our data were generated by means of Monte Carlo simulations for systems of linear size up to $ L=2000 $. Estimates of the exponents for all cluster sets using finite-size scaling were reported, while the appearance of corrections to scaling was monitored by performing fits on intervals $ L_{\text{min}} \le L \le L_{\text{max}} $ with increasing $ L_{\text{min}} $. Overall, where corrections to scaling are not very strong, our results are in  agreement with exact values within error bars and the numerical accuracy allowed us to draw conclusions for the presence of scaling corrections.

The average cluster size is sensitive to the exclusion of certain classes of clusters. Excluding percolating clusters (except clusters percolating in one but not the other direction) or the largest cluster, which is commonly done in numerical studies of random percolation \cite{stauffer_introduction_1994}, leads to strong corrections to scaling. On the other hand, when all clusters are included, minimal corrections to scaling appear. This is also valid when percolating clusters in one but not the other direction are excluded, but mainly because these clusters appear less often.

The percolation strength is less sensitive to the definitions used. Except from the case of $\text{max} \; P_{\text{x and } \overline{\text{y}}}$,  the rest provide reasonable estimates of the involved exponent with rather small corrections. The low probability of clusters that percolate in one but not the other direction, leads to poor statistics resulting in estimates of very low accuracy compared to the other definitions been used.

%

As the average cluster size of the full set $ C $ does not show any maximum as a function of temperature, estimates of pseudo-critical temperatures need to be extracted from other quantities in cases where the transition temperature is not known exactly. A more systematic analysis, allowing one to quantify the corrections to scaling would be an interesting task for future work. The inclusion of additional sets of clusters is also feasible, and could  presumably  lead to optimal sets of definitions for the observables under consideration, with minimal corrections to scaling, thus allowing accurate estimates of the critical exponents.

\ack
We acknowledge the provision of computing time on the parallel computer cluster \emph{Zeus} of Coventry University.


\section*{References}

\bibliography{iopart-num}

\end{document}